\documentclass[aps,pra,superscriptaddress,twocolumn,a4paper,reprint,noeprint]{revtex4-1}
\usepackage[utf8]{inputenc}
\usepackage{graphicx}
\usepackage{amsmath}
\usepackage{amsfonts}
\usepackage{amssymb}
\usepackage{bbold}
\usepackage[usenames,dvipsnames,svgnames,table]{xcolor}
\usepackage[USenglish]{babel}
\usepackage[colorlinks,linkcolor=blue,urlcolor=blue,citecolor=blue]{hyperref}
\usepackage{grffile}

\DeclareMathOperator{\Tr}{Tr\,}

\newcommand{\mytitle}{Stability of Bogoliubov Fermi Surfaces within BCS Theory}

\newcommand{\bdelta}{\mbox{\boldmath$\delta$}}

\begin{document}

\title{\mytitle}

\author{Ankita Bhattacharya}
\email{ankita.bhattacharya@tu-dresden.de}
\affiliation{Institute of Theoretical Physics,
Technische Universit\"at Dresden, 01062 Dresden, Germany}
\author{Carsten Timm}
\email{carsten.timm@tu-dresden.de}
\affiliation{Institute of Theoretical Physics,
Technische Universit\"at Dresden, 01062 Dresden, Germany}
\affiliation{W\"urzburg-Dresden Cluster of Excellence ct.qmat, Technische
Universit\"at Dresden, 01062~Dresden, Germany}


\begin{abstract}
It has recently been realized that the gap nodes of multiband superconductors that break time-reversal symmetry generically take the form of Fermi surfaces of Bogoliubov quasiparticles. However, these Fermi surfaces lead to a nonzero density of states (DOS) at the Fermi energy, which typically disfavors such superconducting states. It has thus not been clear whether they can be stable for reasonable pairing interactions or are in practice preempted by time-reversal-symmetric states with vanishing DOS. In this Letter, we show within BCS theory applied to a paradigmatic model that the time-reversal-symmetry-breaking states are indeed stabilized over broad parameter ranges at weak coupling. Moreover, we introduce a fast method that involves solving the \emph{inverse BCS gap equation}, does not require iteration, does not suffer from convergence problems, and can handle metastable solutions.
\end{abstract}

\maketitle

\textit{Introduction.}---Recent years have witnessed a surge of interest in multiband superconductors (MSCs). In many complex superconductors, multiple bands are close to or cross the Fermi energy \cite{GSZ10,NGR12,NKT16,ong16,nic17,CVF16,agt17,KRO08,ruth94,SMB20,VoG85}. The multiband structure arises from internal degrees of freedom beyond the electron spin, e.g., orbital or basis site. These internal degrees of freedom lead to much richer possibilities for the states of Cooper pairs and hence to superconducting properties that are qualitatively different from the single-band case~\cite{Volovik,ABT17,BAMT18}. Specifically, MSCs enable exotic ``internally an\-i\-so\-tro\-pic'' pairing states \cite{BAMT18}---momentum-in\-de\-pen\-dent \textit{s}-wave pairing can have nontrivial dependence on lattice symmetries.

Moreover, research has been stimulated by the prediction that gap nodes in time-reversal-symmetry-breaking (TRSB) pairing states of MSCs are generically inflated into two-dimensional Fermi surfaces of Bogoliubov quasiparticles (Bogoliubov Fermi surfaces, BFSs)~\cite{Volovik,ABT17,BAMT18,LH20,LBH20,TiB21,Dutta21,Koby21,OnS22}. In single-band models or if interband pairing is neglected, the superconducting gap typically closes at points or lines in momentum space (point or line nodes, respectively) or is everywhere nonzero. BFSs can be viewed as such nodes that are inflated into surfaces by a pseudomagnetic field generated by interband pairing.

The presence of BFSs implies a nonzero DOS at the Fermi energy, which leads to characteristic signatures in, e.g., the specific heat, the penetration depth, and the spin-lattice relaxation rate \cite{LBT20}. MSCs that break time-reversal symmetry (TRS), and are thus potential testbeds for the physics discussed here, include $\mathrm{UPt_3}$ \cite{SGW14,LKL93}, $\mathrm{UBe}_{13}$ and $\mathrm{U}_{1-x}\mathrm{Th}_x\mathrm{Be}_{13}$ \cite{St19}, $\mathrm{Sr_2RuO_4}$ \cite{LFK98,XMB06,GGS21}, $\mathrm{Pr}\mathrm{Os}_4\mathrm{Sb}_{12}$ \cite{Sato03,Maple08}, iron-based superconductors \cite{HK15,HCB16,GWC21}, half-Heusler compounds \cite{KWN18}, and $\mathrm{CeRh}_2\mathrm{As}_2$ \cite{KKK22}. However, the nonzero DOS raises fundamental concerns: Can superconductors with BFSs be energetically stable at all? There are always competing pairing states that do not break TRS and at most have point or line nodes. In such states, electronic spectral weight is pushed away from the Fermi energy compared to superconductors with BFSs, which is expected to lower the free energy. Preliminary estimates in \cite{ABT17} and BCS results from \cite{MTB19}, where the main interest was in the role of spin-orbit coupling, indicate that BFSs could be stable under realistic conditions. A recently suggested orbital-antisymmetric spin-triplet pairing state of $\mathrm{Sr_2RuO_4}$ also features BFSs~\cite{SMB20}.

In this Letter, we study a prototypical model \cite{BWW16,ABT17,BAMT18,RGF19} of cubic superconductors describing electrons with an effective angular momentum $j=3/2$ within BCS theory \cite{BCS}. We restrict ourselves to local pairing. Nonlocal pairing allows for additional symmetries of pairing states but does not introduce fundamentally new aspects~\cite{TiB21}.

The standard approaches to obtain the pairing amplitudes within BCS theory are to either solve the gap equation self-consistently or to minimize the free energy. These methods are computationally expensive since they require iteration, and they may fail to converge. We adopt a different strategy, namely to solve the \emph{inverse gap equation}, which avoids these problems. We find that TRSB states with BFSs are indeed favored over broad parameter ranges for not too strong pairing interactions.

\textit{Model.}---We consider an effective spin-$3/2$ model with point group $O_h$ \cite{BWW16,ABT17,BAMT18,RGF19}. The Hilbert space of internal degrees of freedom is thus four dimensional. The effective spin $3/2$ emerges due to the presence of strong spin-orbit coupling and provides the natural description of electrons close to a fourfold $\Gamma_{8}^+$ ($F_{3/2,g}$) band-touching point. The concept of BFSs is not limited to $j=3/2$ systems; the multiband nature may also arise due to other internal degrees of freedom~\cite{BAMT18,TiB21}.

We expand the normal-state Hamiltonian into a basis of Hermitian $4\times 4$ matrices $h_n$ as
\begin{equation}
H_N(\mathbf{k}) = \sum_{n=0}^5 c_n(\mathbf{k})\, h_n .
\label{2.HNseries.3}
\end{equation}
The functions $c_n(\mathbf{k})$ are periodic basis functions of irreducible representations (irreps) of $O_h$ that transform in the same way as the corresponding matrices $h_n$ so that $H_N(\mathbf{k})$ is invariant under $O_h$. The matrices $h_n$ can be chosen such that $h_1,\ldots,h_5$ anticommute pairwise and $h_0$, the $4\times 4$ identity matrix, commutes with any matrix \cite{TiB21}. The basis matrices $h_n$ and the functions $c_n(\mathbf{k})$ are listed in the Supplemental Material (SM) \cite{SM}. The normal-state eigenenergies are $\xi_{\mathbf{k}\pm} = c_0(\mathbf{k}) \pm \sqrt{c_1^2(\mathbf{k}) + \ldots + c_5^2(\mathbf{k})}$.

Even-parity local pairing allows for six possible pairing channels, which belong to the one-dimensional irrep $A_{1g}$, the two-di\-men\-sio\-nal irrep $E_g$, and the three-di\-men\-sio\-nal irrep $T_{2g}$ \cite{ABT17,BAMT18,RGF19}. After mean-field decoupling, the pairing interaction reads as~\cite{BWW16,ABT17,BAMT18}
\begin{align}
H_\mathrm{BCS}^\mathrm{int}
  &= \frac{1}{2}\, \sum_{\mathbf{k}} \sum_{n=0}^5 \Big[ \Delta^*_{n}\,
    c_{-\mathbf{k}}^T \, (h_n U_T)^\dagger\, c_{\mathbf{k}} \nonumber \\
&\quad{}+ \Delta_{n}\, c_{\mathbf{k}}^\dagger\, h_n U_T \,{c_{-\mathbf{k}}^{\dagger T}}
  \Big]
  + \frac{N}{2} \sum_{n=0}^5 \frac{|\Delta_{n}|^2}{V_{n}} .
\end{align}
Here, $N$ is the number of unit cells, and for attractive pairing interactions we take $V_n>0$. The pairing amplitudes are given by
\begin{equation}
\Delta_n \equiv \Delta_n^1 + i \Delta_n^2
  = -\frac{V_n}{N} \sum_{\mathbf{k}} \langle c^T_{-\mathbf{k}}\, (h_n U_T)^\dagger \,
     c_{\mathbf{k}}\rangle,
\end{equation}
where $U_T = \exp(i J_y \pi)$ is the unitary part of the time-reversal operator. Neglecting an irrelevant constant, the full BCS Hamiltonian can then be written as
\begin{equation}
H_\mathrm{BCS} = \frac{1}{2} \sum_\mathbf{k} \Psi^\dagger_\mathbf{k} \mathcal{H}(\mathbf{k})
  \Psi_\mathbf{k}
  + \frac{N}{2} \sum_{n=0}^5 \frac{|\Delta_{n}|^2}{V_{n}}
\end{equation}
in terms of the Nambu spinor $\Psi_\mathbf{k} = ( c_\mathbf{k}, c^\dagger_{-\mathbf{k}} )^T$ and the Bogoliubov--de Gennes (BdG) Hamiltonian
\begin{equation}
\mathcal{H}(\mathbf{k}) = \begin{pmatrix}
    H_N(\mathbf{k}) & \hat\Delta \\
    \hat\Delta^\dagger & -H_N^T(-\mathbf{k})
  \end{pmatrix} ,
\end{equation}
where $\hat\Delta = \sum_{n=0}^5\, (\Delta_n^1 + i \Delta_n^2)\, h_n U_T$.

\textit{Inverse gap equation.}---The free energy per unit cell resulting from $H_\mathrm{BCS}$ reads as~\cite{DaS07,SMB20}
\begin{equation}
F = -\frac{k_BT}{2N} \sum_\mathbf{k} \sum_{i=1}^8 \ln\!\big(1 + e^{-\beta E_{\mathbf{k},i}}\big)
  + \frac{1}{2} \sum_{n=0}^5\, \sum_{\alpha=1}^2 \frac{(\Delta_n^\alpha)^2}{V_n^\alpha} ,
\end{equation}
where $E_{\mathbf{k},i}$ are the eigenvalues of the BdG Hamiltonian, $i$ denotes the band, and $\beta=1/k_BT$ is the inverse temperature. $E_{\mathbf{k},i}$ can be obtained in closed form, as discussed in the SM \cite{SM}. We temporarily allow the interaction strength $V_n^\alpha$ to depend on the index $\alpha$ referring to the real and imaginary parts of the pairing amplitudes.

Equating the derivative of the free energy with respect to $\Delta_n^\alpha$ to zero, we obtain the BCS gap equation
\begin{equation}
\Delta_{n}^\alpha = \frac{V_n^\alpha}{2N} \sum_\mathbf{k} \sum_{i=1}^4
  \tanh{\frac{\beta E_{\mathbf{k},i}}{2}}\,
  \frac{\partial E_{\mathbf{k},i}}{\partial \Delta_n^\alpha}
  \equiv V_n^\alpha f_{n}^\alpha(\mathbf{\Delta}) ,
\label{gapEqn}
\end{equation}
where we have used that the spectrum at fixed $\mathbf{k}$ is symmetric. $\mathbf{\Delta} \equiv (\Delta^1_0,\ldots,\Delta^1_5;\Delta^2_0,\ldots,\Delta^2_5) \in \mathbb{R}^{12}$ is the vector of all order parameters. $f_{n}^\alpha(\mathbf{\Delta})$ has to be of at least first order in $\Delta_{n}^\alpha$ since otherwise the normal-state solution $\mathbf{\Delta}=0$ would not exist \cite{SM}. Hence, we can write $f_{n}^\alpha(\mathbf{\Delta})\equiv \Delta_{n}^\alpha\, g_{n}^\alpha(\mathbf{\Delta})$. For $\Delta_{n}^\alpha\neq 0$, solving Eq.\ (\ref{gapEqn}) for $V_n^\alpha$ yields the inverse gap equation
\begin{equation}
V_n^\alpha = \frac{1}{\displaystyle g_n^\alpha(\mathbf{\Delta})} .
\label{invgap2}
\end{equation}
This equation describes the interaction strength that is necessary to obtain a given $\mathbf{\Delta}$. If $g_{n}^\alpha(\mathbf{\Delta})$ vanishes and $V_n^\alpha$ thus diverges the given $\mathbf{\Delta}$ cannot be stabilized by any value of the interaction. On the other hand, for $\Delta_{n}^\alpha=0$, Eq.\ (\ref{gapEqn}) becomes tautological ($0=0$) and there is no constraint on $V_n^\alpha$---a state without pairing in the $\Delta_{n}^\alpha$ channel is compatible with any value of $V_n^\alpha$ (the state might not be the global minimum, though).

We first discuss the case without any symmetries. In this case, the parameter space is spanned by $6\times 2 = 12$ independent coupling constants $V_n^\alpha$. In principle, one can scan the $12$-component order parameter $\mathbf{\Delta}$ over the relevant ranges and obtain the $12$ coupling constants $\mathbf{V} \equiv (V_0^1,\ldots,V_5^1;V_0^2,\ldots,V_5^2)$ from Eq.\ (\ref{invgap2}). Then the mapping is inverted to obtain $\mathbf{V} \mapsto \mathbf{\Delta}$. This is generically possible at least locally since its domain and codomain have the same dimension. If there are multiple solutions for a $\mathbf{V}$, then the solution with the minimum free energy is stable.

If $s$ order parameters $\Delta_n^\alpha$ vanish, this corresponds to a ($12-s$)-dimensional subspace of the space of $\mathbf{\Delta}$. The inverse gap equations map this subspace into a codomain of generically the same dimension $12-s$. The $s$ coupling constants belonging to the vanishing $\Delta_n^\alpha$ are arbitrary. Hence, restricting any number of order parameters to zero does not reduce the dimension of the space of allowed coupling constants. Indeed, in the normal state, all $\Delta^\alpha_n$ vanish and all $V^\alpha_n$ are unconstrained.

Symmetries reduce the number of independent coupling constants. Due to the global $\mathrm{U}(1)$ symmetry, $V_{n}^\alpha$ does not depend on $\alpha$ and we thus drop the superscript again. The presence of cubic symmetry further restricts the number of independent $V_n$. For local pairing, the six components of the pairing matrix $\hat\Delta$ transform according to the three irreps $A_{1g}$, $T_{2g}$, and $E_g$ \cite{BWW16,BAMT18,TiB21}. The coupling constants belonging to the same irrep must be equal. Hence, we can organize them as
\begin{align}
    A_{1g}&:\quad V_A \equiv V_0, \\
    T_{2g}&:\quad V_T \equiv V_1 = V_2 = V_3, \\
    E_{g} &:\quad V_E \equiv V_4 = V_5.
\end{align}
Note that the inverse gap equations still generically map the 12-dimensional space of $\mathbf{\Delta}$ onto a 12-dimensional image of couplings $\mathbf{V}$ but now only a three-dimensional submanifold is allowed. Random points $\mathbf{\Delta}$ map onto unphysical $\mathbf{V}$ with probability one. We thus need to restrict the values of $\mathbf{\Delta}$. Landau theory is very helpful to identify likely relations between amplitudes $\Delta_n^\alpha$ belonging to the same irrep \cite{SiU91,BWW16,BAMT18}. For example, the only fundamentally different patterns of pairing amplitudes expected for the $E_g$ channels are $(\Delta_4,\Delta_5) \propto (1,0)$, $(0,1)$, and $(1,i)$.

For pure-irrep pairing, the $\Delta_n^\alpha$ belonging to all but one irrep vanish. The couplings $V_n$ for all but this single irrep are then unconstrained. We solve the inverse gap equation for each of the expected patterns of pairing amplitudes for this irrep. These are one-to-one mappings and thus easy to invert. Mixed-irrep pairing states can also be treated: Let there be $n_\Gamma$ irreps with nonzero pairing. Free-energy arguments provide a set of plausible patterns, now depending on $n_\Gamma$ independent amplitudes, one for each irrep involved. The inverse gap equation provides an $n_\Gamma$-to-$n_\Gamma$ mapping from pairing amplitudes to couplings, which can generically be inverted. The free energy of the resulting pure-irrep and mixed-irrep solutions has to be compared to find the stable state.

The main advantages of the inverse gap equation are that it does not require iteration to reach selfconsistency and thus avoids convergence problems and is much faster. Moreover, it easily deals with metastable and unstable branches in the vicinity of first-order phase transitions.

We emphasize that high precision in the integration over momentum space is essential for reaching the weak-coupling regime. This is numerically difficult because the weak-coupling behavior relies on a term proportional to $\Delta^2 \ln (\Delta/\Lambda)$ in the free (internal) energy at $T=0$, which has to be separated from a $\Delta^2$ contribution. We observe that summing over a momentum-space mesh becomes forbiddingly slow for three-dimensional systems. Instead, we have obtained good results using adaptive integration. Details are discussed in the SM~\cite{SM}. The high precision required for the three-dimensional momentum integration makes iterative methods prohibitively costly.

\textit{Results.}---In this Letter, we restrict ourselves to zero temperature. We first discuss $E_g$ pairing states, which are described by the two-component order parameter $(\Delta_4,\Delta_5) \equiv \Delta\, (\delta_{x^2-y^2},\delta_{3z^2-r^2}) \equiv \Delta\, \bdelta$ \cite{BAMT18}. The pairing matrix reads as
\begin{equation}
\hat\Delta = \Delta \left( \delta_{x^2-y^2}\, h_4 U_T + \delta_{3z^2-r^2}\, h_5 U_T \right) ,
\end{equation}
where the basis matrices $h_4$ and $h_5$ transform like $x^2-y^2$ and $3z^2-r^2$, respectively,  under $O_h$ \cite{SM}. As noted above, potentially stable pairing states are the time-reversal-symmetric states with the patterns $\bdelta = (1,0)$ and $(0,1)$ and the TRSB state with $\bdelta = (1,i)/\sqrt{2}$~\cite{BWW16,BAMT18}.

\begin{figure}
\raisebox{1ex}{(a)}\includegraphics[width=0.85\columnwidth]{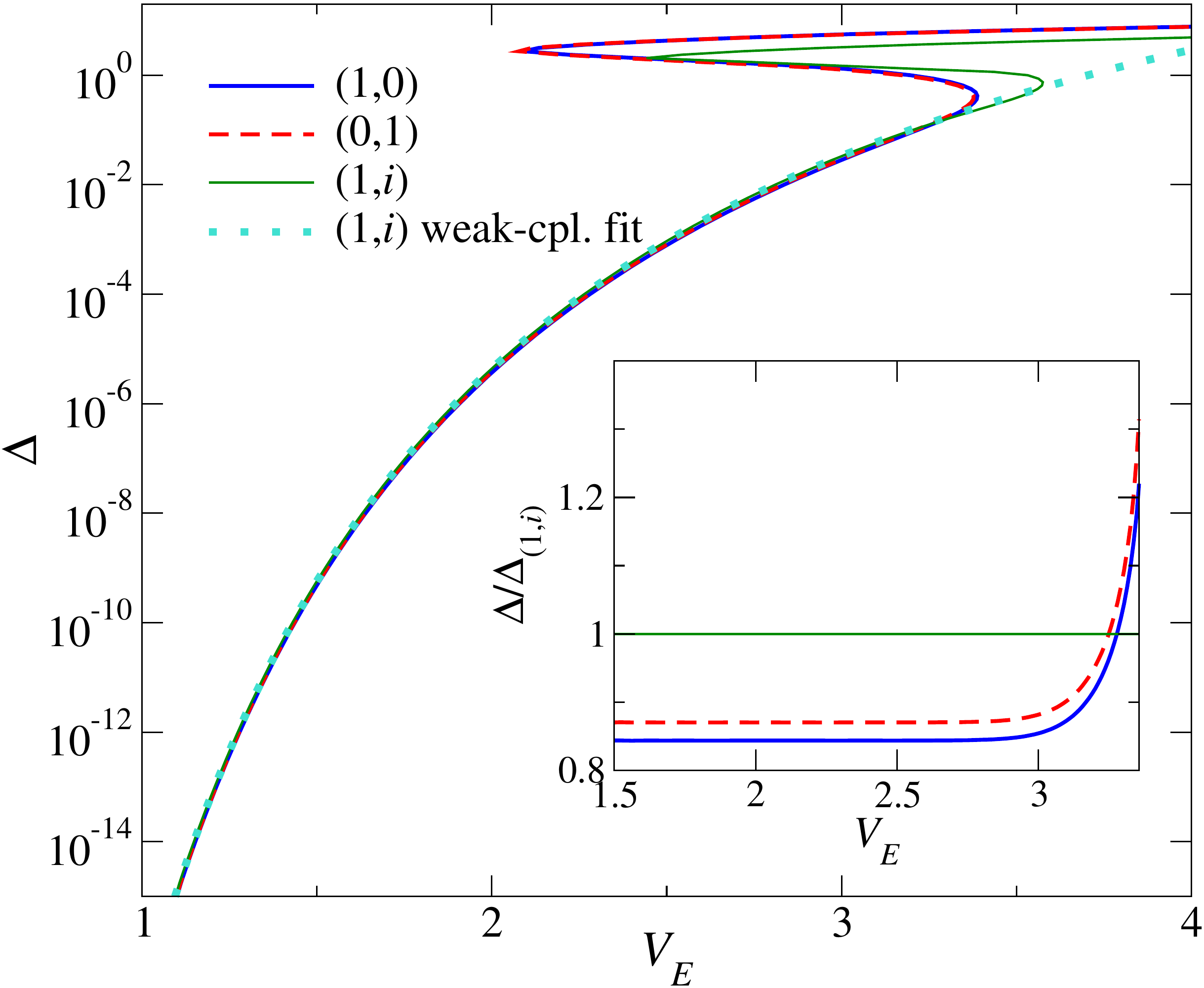}\\
\raisebox{1ex}{(b)}\includegraphics[width=0.85\columnwidth]{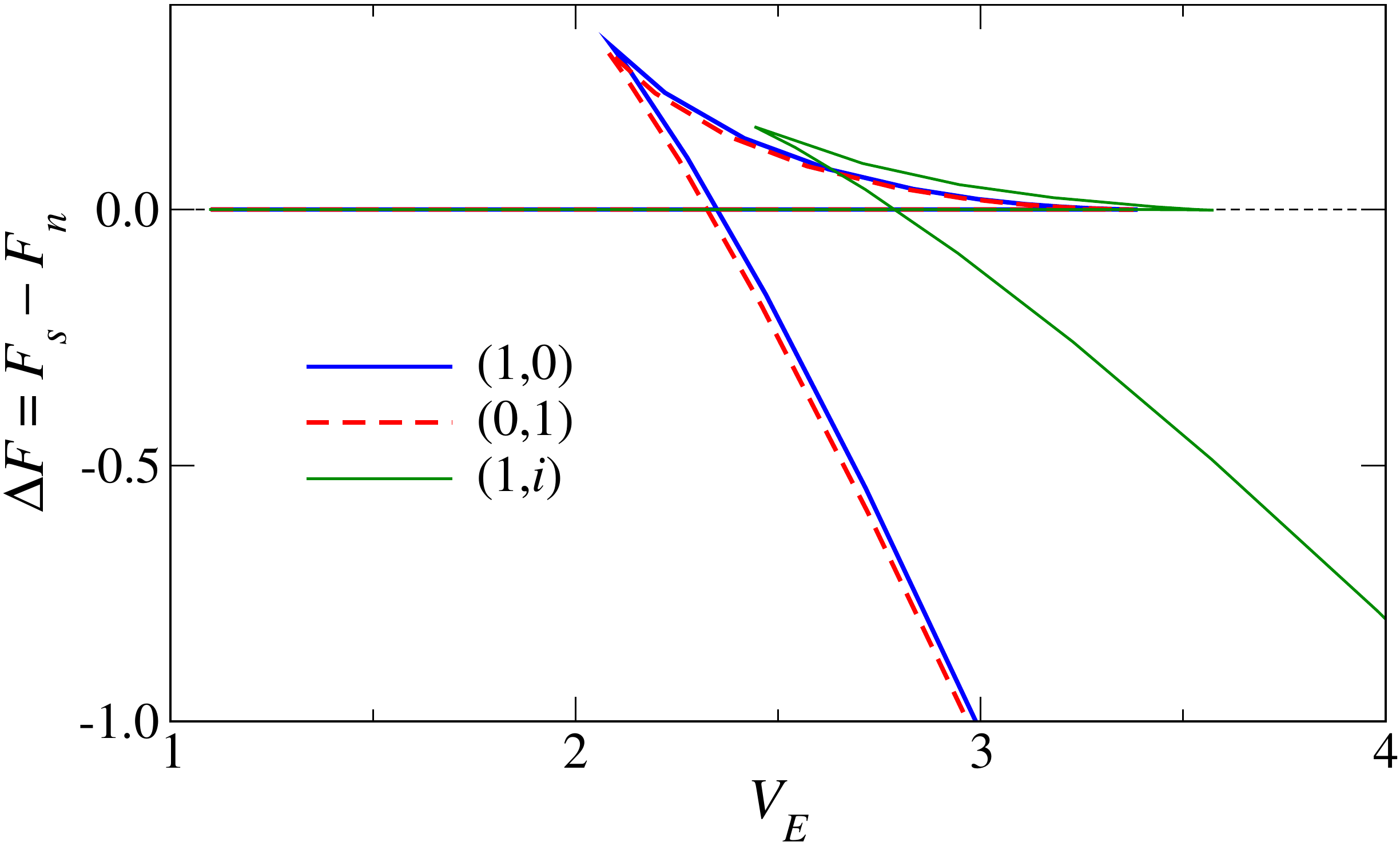}\\[0.5ex]
\raisebox{1ex}{(c)}\includegraphics[width=0.85\columnwidth]{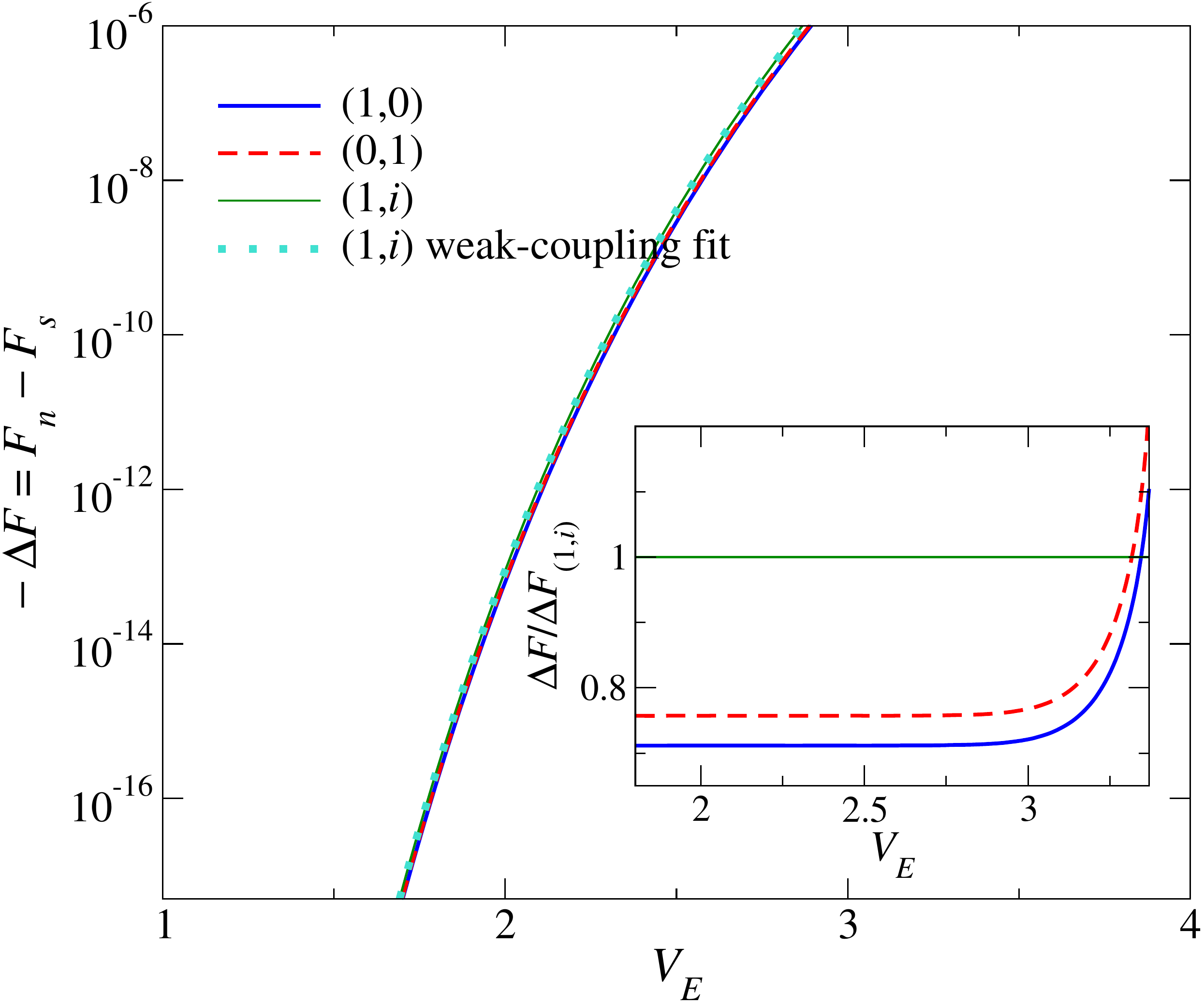}
\caption{(a) Pairing amplitudes $\Delta$ as functions of the coupling strength $V_E$ at $T=0$ for the $E_g$ pairing states $(1,0)$, $(0,1)$, and $(1,i)$. At small $V_E$, $\Delta(V_E)$ shows weak-coupling scaling, $\ln\Delta \sim A - B/V_E$. The dotted curve is a fit of this scaling form to the numerical results for the $(1,i)$ state at small $V_E$.
The inset shows the ratios of $\Delta$ for the three states to $\Delta$ for the $(1,i)$ state.
(b) Free-energy differences $\Delta F = F_s-F_n$ between the superconducting and normal states as functions of the coupling strength $V_E$ at $T=0$. The state with minimal (most negative) $\Delta F$ is stable.
(c) The same on a logarithmic scale, for the weak-coupling branches. Note that $-\Delta F>0$ is plotted. At small $V_E$, we observe the weak-coupling scaling $\ln(-\Delta F) \sim A' - B'/V_E$. The dotted curve shows the weak-coupling scaling function for $\Delta F$ obtained from the fit in (a). The inset shows the ratios of $\Delta F$ for the three states to $\Delta F$ for the $(1,i)$ state. Numerical parameters are given in the SM~\cite{SM}.}
\label{Fig1}
\end{figure}

Figure \ref{Fig1}(a) shows the pairing amplitudes $\Delta$ as functions of the coupling strength $V_E$ for the three $\bdelta$. At weak coupling, $\Delta(V_E)$ shows the expected weak-coupling scaling $\ln\Delta \sim A - B/V_E$ with constants $A$, $B$ \cite{SM}. For large coupling strength, $\Delta$ exhibits a pronounced S shape, which is characteristic of a first-order phase transition. Correspondingly, the free-energy differences between superconducting and normal states, $\Delta F \equiv F_s-F_n$, exhibit a swallowtail feature as a function of $V_E$, as shown in Fig.\ \ref{Fig1}(b). A similar self-crossing of the free energy was observed for superconducting states with finite-momentum Cooper pairs by Fulde and Ferrell \cite{FF64}, where the control parameter was an exchange field.

In the range of $V_E$ with three solutions, the one with the intermediate value of $\Delta$ and the highest value of the free energy corresponds to a maximum separating two (meta-) stable states. The S shape thus means that for increasing coupling strength a second solution with much larger pairing amplitude $\Delta$ appears, which initially is metastable. This solution can be attributed to interband pairing: The values of $\Delta$ for this branch are comparable to the energy difference between the normal-state bands. Hence, the superconducting pairing can take advantage of the additional DOS in these bands.

If there were a single $E_g$ pairing state our results would predict a first-order transition without a change in symmetry similar to a liquid-gas transition. However, there are three distinct pairing states controlled by the same $V_E$. For weak coupling, the TRSB $(1,i)$ state is favored, see the inset in Fig.\ \ref{Fig1}(c). Note that the favored state corresponds to the highest value since $\Delta F_{(1,i)}<0$. However, at strong coupling, the $(1,i)$ state becomes strongly disfavored and there is a small preference for the $(0,1)$ state over the $(1,0)$ state. The $(1,0)$ state has two crossing line nodes ($x^2-y^2$ symmetry) \cite{RGF19}, which leads to a higher DOS close to the Fermi energy and is thus energetically unfavorable \cite{SiU91}, compared to the $(0,1)$ state with non-crossing line nodes ($3z^2-r^2$ symmetry).

Figure \ref{Fig1}(c) shows the free-energy gain on a logarithmic scale at weak coupling. The free-energy gain follows the expected behavior $\ln(-\Delta F) \sim A' - B'/V_E$, where the parameters are not free but determined by the scaling form of $\Delta$ \cite{SM}. We again see that the TRSB state is favored in the weak-coupling limit. The inset in Fig.\ \ref{Fig1}(c) shows that the energy separation between the three states is sizable on the relevant energy scale. We emphasize that the smaller values of $\Delta$ and $\Delta F$ shown in Fig.\ \ref{Fig1} would be experimentally out of reach. The main reason for plotting them here is to test the solution of the BCS gap equation---the numerical method remains viable deep into the weak-coupling regime.

The weak-coupling results can be understood following Ref.\ \cite{SiU91} since the TRSB $(1,i)$ state has point nodes in the limit of small $V_E$ and thus has a lower DOS close to the Fermi energy than the $(1,0)$ and $(0,1)$ states, which have line nodes. With increasing coupling $V_E$, the point nodes are inflated into BFSs, which have larger DOS and are disfavored. The line nodes of the time-reversal-symmetric states are not inflated. Thus, at strong coupling, the TRSB state is destabilized.

We briefly comment on the $T_{2g}$ pairing states. Similar to $E_g$ pairing, we find a first-order transition from a TRSB state at weak coupling to the time-reversal-symmetric state at strong coupling. Details are shown in the SM~\cite{SM}.

\begin{figure}
\includegraphics[width=0.85\columnwidth]{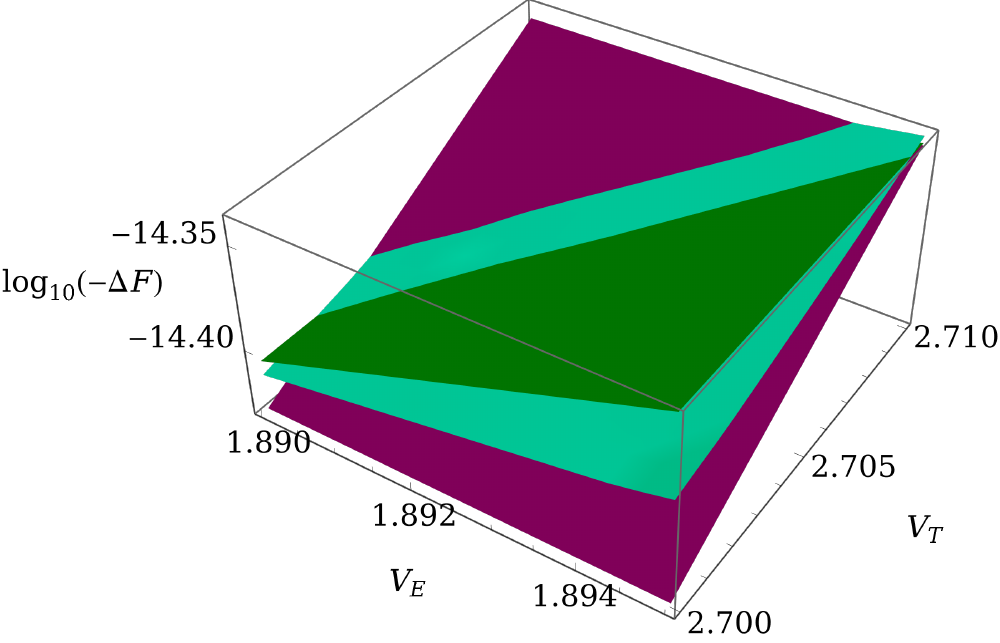}\\
\hspace{4.5em}\includegraphics[width=0.45\columnwidth]{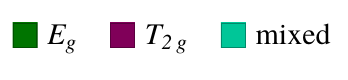}
\caption{Free-energy differences $\Delta F = F_s - F_n$ between the superconducting and normal states as functions of the coupling strenghts $V_E$ and $V_T$. A pure $E_g$ state, a pure $T_{2g}$ state, and a mixed-irrep state are compared, see text. Note that $-\Delta F>0$ is plotted on a logarithmic scale so that the largest value corresponds to the stable state.
}
\label{fig:mixed}
\end{figure}

So far, we have discussed pairing states that belong to a single irrep. However, mixed-irrep pairing states are possible when two pure-irrep states are nearly degenerate. We consider the mixed-irrep state $\Delta_{xy}^{T_{2g}} + i \Delta_{x^2-y^2}^{E_g}$, where the subscript describes the symmetry of the order parameter. This pairing state has only two point nodes in the limit $\Delta \to 0$ \cite{RGF19} and should thus be favored over all other mixed-irrep states, which have line nodes or more point nodes. We compare the free-energy gain as a function of the couplings $V_E$ and $V_T$ for this state to the pure $E_g$ state with $\bdelta = (1,i)/\sqrt{2}$ and the pure $T_{2g}$ state with $\bdelta = (1,i,0)/\sqrt{2}$ in the weak-coupling regime in Fig.\ \ref{fig:mixed}. As discussed above, the result for the pure $E_g$ state does not depend on the interaction in the $T_{2g}$ channel and vice versa. We find that the mixed-irrep state is stabilized in a narrow region of coupling constants.
This result is reasonable since the mixed-irrep state with two (inflated) point nodes is energetically favored over the $E_g$ state with eight point nodes and the $T_{2g}$ state with two point nodes and one line node \cite{BAMT18} if the couplings are fine tuned close to degeneracy.

\textit{Conclusions.}---We have performed a BCS study of a paradigmatic centrosymmetric spin-$3/2$ model to analyze the stability of TRSB superconducting states with BFSs. At weak coupling, such TRSB states are indeed stable. For increasing pairing interactions, TRSB states become disfavored compared to time-reversal-symmetric states due to the increasing DOS at the Fermi energy resulting from the growing BFSs. This eventually leads to a first-order transition to a time-reversal-symmetric state at strong coupling. If there were only a single possible pairing state, our results would predict a liquid-gas-like first-order transition without a change in symmetry.

We have proposed an alternative approach to solve the BCS gap equations. The main idea is to solve the inverse equations to obtain the coupling strengths as functions of the pairing amplitudes and to invert the resulting mapping. If there is only a single coupling and a single amplitude, this inversion is a trivial interchange of axes but the method works beyond this case. The inverse gap equation avoids iterations and the associated convergence problems. It can also treat metastable and unstable branches as well as the resulting first-order transitions, which is essential for this work.

The approach has been derived for general temperatures and illustrated for the limit $T\to 0$. The application at $T>0$ should in fact be numerically more benign since the weak-coupling scaling is cut off by $T$. The interesting questions of the relative stability of states with and without BFSs at $T>0$ and the fate of the metastable solutions are left for future work.

\medskip
The authors thank Philip M. R. Brydon and Julia M. Link for useful discussions. Financial support by the Deutsche Forschungsgemeinschaft, in part through Collaborative Research Center SFB 1143, project A04, project id 247310070, and the W\"urzburg-Dresden Cluster of Excellence ct.qmat, EXC 2147, project id 390858490, is gratefully acknowledged.

\clearpage
\onecolumngrid

\setcounter{equation}{0}
\setcounter{figure}{0}
\setcounter{page}{1}

\renewcommand{\theequation}{S\arabic{equation}}
\renewcommand{\thefigure}{S\arabic{figure}}

\begin{center}
{\large\textbf{Supplemental Material for}}\\[1ex]
{\large\textbf{\mytitle}}\\[1.5ex]
Ankita Bhattacharya and Carsten Timm
\end{center}
\section{Basis matrices and details of the model}
\label{SM.sec.model}

In this section, we summarize the basis matrices that appear in both the normal-state Hamiltonian $H_N(\mathbf{k})$ and in the superconducting pairing matrix $\hat\Delta$. We also present details of the model Hamiltonian. In terms of the standard spin-$3/2$ matrices
\begin{align}
J_x &= \frac{1}{2} \begin{pmatrix}
  0 & \sqrt{3} & 0 & 0 \\
  \sqrt{3} & 0 & 2 & 0 \\
  0 & 2 & 0 & \sqrt{3} \\
  0 & 0 & \sqrt{3} & 0
  \end{pmatrix} , \\
J_y &= \frac{i}{2} \begin{pmatrix}
  0 & -\sqrt{3} & 0 & 0 \\
  \sqrt{3} & 0 & -2 & 0 \\
  0 & 2 & 0 & -\sqrt{3} \\
  0 & 0 & \sqrt{3} & 0
  \end{pmatrix} , \\
J_z &= \frac{1}{2} \begin{pmatrix}
  3 & 0 & 0 & 0 \\
  0 & 1 & 0 & 0 \\
  0 & 0 & -1 & 0 \\
  0 & 0 & 0 & -3
  \end{pmatrix} ,
\end{align}
the basis matrices are given by~\cite{SM.ABT17,SM.SRV17,SM.BAMT18}
\begin{align}
h_0 &= \mathbb{1} , \\
h_1 &= \frac{J_y J_z + J_z J_y}{\sqrt{3}} , \\
h_2 &= \frac{J_z J_x + J_x J_z}{\sqrt{3}} , \\
h_3 &= \frac{J_x J_y + J_y J_x}{\sqrt{3}} , \\
h_4 &= \frac{J_x^2 - J_y^2}{\sqrt{3}} , \\
h_5 &= \frac{2J_z^2 - J_x^2 - J_y^2}{3} ,
\end{align}
which are orthonormalized so that $\Tr h_m h_n = 4\delta_{mn}$. The matrices $h_1$ to $h_5$ transform under point-group operations like the basis functions $yz$, $zx$, $xy$, $x^2-y^2$, and $3z^2-r^2$. $h_1$ to $h_3$ are irreducible tensor operators belonging to the irrep $T_{2g}$, while $h_4$ and $h_5$ are irreducible tensor operators belonging to $E_g$.

The normal-state Hamiltonian can be expanded as
\begin{equation}
H_N(\mathbf{k}) = \sum_{n=0}^5 c_n(\mathbf{k})\, h_n .
\end{equation}
Under point-group operations, the functions $c_n(\mathbf{k})$ must transform in the same way as the $h_n$ so that the full Hamiltonian transforms trivially. The $c_n(\mathbf{k})$ must also respect the periodicity of the reciprocal lattice. We assume a face-centered-cubic lattic, which is appropriate for example for an effective model of a pyrochlore \cite{SM.KBT22} and for Heusler compounds if antisymmetric spin-orbit coupling is negligible \cite{SM.TSA17}. The functions $c_n(\mathbf{k})$ are chosen as
\begin{align}
c_0(\mathbf{k}) &= (-4 t_1 - 5 t_2) (\cos k_x \cos k_y + \cos k_y \cos k_z
  + \cos k_z \cos k_x) - \mu , \\
c_1(\mathbf{k}) &= 4\sqrt{3}\, t_3 \sin k_y \sin k_z , \\
c_2(\mathbf{k}) &= 4\sqrt{3}\, t_3 \sin k_z \sin k_x , \\
c_3(\mathbf{k}) &= 4\sqrt{3}\, t_3 \sin k_x \sin k_y , \\
c_4(\mathbf{k}) &= -2\sqrt{3}\, t_2 (\cos k_y \cos k_z - \cos k_z \cos k_y) , \\
c_5(\mathbf{k}) &= 2t_2 (\cos k_y \cos k_z + \cos k_z \cos k_x - 2 \cos k_x \cos k_y) .
\end{align}
Including higher-order trigonometric functions, corresponding to longer-range hopping in real space, would not affect the qualitative results. For the numerical calculations, we set $t_1=-0.918\,\mathrm{eV}$, $t_2=-0.760\,\mathrm{eV}$, $t_3=-0.253\,\mathrm{eV}$, and $\mu = -0.88\,\mathrm{eV}$. These values originally come from a tight-binding fit to density-functional results for the band structure of the half-Heusler compound YPtBi, neglecting the inversion-symmetry-breaking antisymmetric spin-orbit coupling, see Ref.~\cite{SM.TSA17}.

\section{Analytical expressions}

In the following, we present the secular equations for the eigenenergies of the BdG Hamiltonian. We also give useful closed expressions for the coefficients of these equations and for derivatives of eigenenergies with respect to the pairing amplitudes.

\subsection{Closed form of eigenenergies}

The eigenvalues of the BdG Hamiltonian
\begin{equation}
\mathcal{H}(\mathbf{k}) = \begin{pmatrix}
    H_N(\mathbf{k}) & \hat\Delta \\
    \hat\Delta^\dagger & -H_N^T(-\mathbf{k})
  \end{pmatrix}
\end{equation}
can be obtained in analytical form.
The secular equation for eigenvalues $E$ of $\mathcal{H}$ is of order eight but the presence of both inversion symmetry and charge-conjugation symmetry guarantees that the solutions come in pairs of opposite sign. This allows us to reduce the secular equation to two quartic equations
\begin{equation}
E^4 + p E^2 \pm q E + r = 0 .
\label{SM.quartic2}
\end{equation}
It is sufficient to solve one of them since the solutions of the other one are simply the negative of the solutions of the first. We consider the first one,
\begin{equation}
E^4 + p E^2 + q E + r = 0 .
\label{SM.quartic3}
\end{equation}
Since the equation is of depressed form (there is no $E^3$ term) the solutions satisfy $E_1 + E_2 + E_3 + E_4 = 0$. To give the analytical expressions for the coefficients $p$, $q$, and $r$, we define the five-component vectors
\begin{align}
\vec{c} &\equiv (c_1,c_2,c_3,c_4,c_5) ,
\label{SM.def.c} \\
\vec{\Delta}^1 &\equiv (\Delta_1^1,\Delta_2^1,\Delta_3^1,\Delta_4^1,\Delta_5^1) , \\
\vec{\Delta}^2 &\equiv (\Delta_1^2,\Delta_2^2,\Delta_3^2,\Delta_4^2,\Delta_5^2) .
\end{align}
In Eq.\ (\ref{SM.def.c}), we have suppressed the momentum argument. We further define the Gram matrix of these vectors~\cite{SM.HJ12},
\begin{equation}
M \equiv \begin{pmatrix}
  \vec{c}\cdot\vec{c} & \vec{c}\cdot\vec{\Delta}^1 & \vec{c}\cdot\vec{\Delta}^2  \\
  \vec{\Delta}^1\cdot\vec{c} & \vec{\Delta}^1\cdot\vec{\Delta}^1 &
    \vec{\Delta}^1\cdot\vec{\Delta}^2 \\
  \vec{\Delta}^2\cdot\vec{c} & \vec{\Delta}^2\cdot\vec{\Delta}^1 &
   \vec{\Delta}^2\cdot\vec{\Delta}^2
  \end{pmatrix} .
\end{equation}
Being a Gram matrix, $M$ is real and symmetric and also positive semidefinite. The following expressions have been derived with the help of Mathematica \cite{SM.mathematica}, making use of the $\mathrm{SO}(5)$ invariance of the eigenenergies under simultaneous rotations of the vectors $\vec{c}$, $\vec{\Delta}^1$, and $\vec{\Delta}^2$, which follows from the expansion of $\mathcal{H}(\mathbf{k})$ into 18 basis matrices and their commutation relations described in Ref.~\cite{SM.TiB21}.

The first coefficient $p$ reads as
\begin{equation}
p = -2 \left[\sum_{n=0}^5 c_n^2+ \sum_{n=0}^5 (\Delta_n^1)^2
  + \sum_{n=0}^5  (\Delta_n^2)^2\right]
  = -2 \left[c_0^2 +  (\Delta_0^1)^2 + (\Delta_0^2)^2\right]- 2 \, \Tr M .
\label{SM.p}
\end{equation}
This coefficient is clearly nonzero and negative whenever $\mathcal{H}$ is not the null matrix. The characteristic energy scale of the BdG Hamiltonian is $\sqrt{-p/2}$.

The second coefficient $q$ is more interesting. If $q$ vanishes the quartic equations (\ref{SM.quartic2}) become biquadratic and have the solutions $E_1$, $-E_1$, $E_2$, $-E_2$. Since the two quartic equations become identical all eigenenergies are at least twofold degenerate. Since they must be twofold degenerate in the presence of inversion symmetry and TRS by the Kramers theorem these symmetries imply $q=0$. The reverse statement does not hold. The coefficient $q$ is given by
\begin{align}
q = 8\, \sqrt{\det M}
&= 8\, \Big[(\vec{c}\cdot\vec{c})(\vec{\Delta}^1\cdot \vec{\Delta}^1)
    (\vec{\Delta}^2\cdot \vec{\Delta}^2)]
  + 2 (\vec{c}\cdot\vec{\Delta}^1)(\vec{\Delta}^1\cdot \vec{\Delta}^2)
    (\vec{\Delta}^2\cdot \vec{c})
  - (\vec{c}\cdot\vec{c})(\vec{\Delta}^1\cdot \vec{\Delta}^2)^2 \\ \nonumber
&\quad{}- (\vec{\Delta}^1\cdot \vec{\Delta}^1)(\vec{\Delta}^2\cdot \vec{c})^2
  - (\vec{\Delta}^2\cdot \vec{\Delta}^2)(\vec{c}\cdot \vec{\Delta}^1)^2\Big]^{1/2} .
\label{SM.q}
\end{align}
Nonzero $q$ implies a nonzero Gram matrix $M$ and thus that the three five-component vectors $\vec{c}$, $\vec{\Delta}^1$, and $\vec{\Delta}^2$ are linearly independent. Therefore, twofold degeneracy of the eigenvalues requires $\vec{c}$, $\vec{\Delta}^1$, and $\vec{\Delta}^2$ to be coplanar (or zero). We note that $q$ does not depend on the ``singlet'' components $c_0$, $\Delta_0^1$, and $\Delta_0^2$.

Using the generalized ``Minkowski'' product
\begin{align}
\langle A, B \rangle \equiv A_0 B_0- \vec{A}\cdot\vec{B} ,
\end{align}
the third coefficient $r$ can be written as
\begin{align}
r &= \langle c, c \rangle^2 + \langle \Delta^1, \Delta^1 \rangle^2
  + \langle \Delta^2, \Delta^2 \rangle^2
  + 4\, ( \langle c, \Delta^1 \rangle^2 + \langle \Delta^1, \Delta^2 \rangle^2
  + \langle \Delta^2, c\rangle^2) \nonumber \\
&\quad{}- 2\, ( \langle c,c \rangle \langle \Delta^1, \Delta^1 \rangle
  + \langle \Delta^1, \Delta^1 \rangle \langle \Delta^2, \Delta^2 \rangle
  + \langle \Delta^2, \Delta^2 \rangle \langle c,c \rangle) .
\label{SM.Pfrl}
\end{align}
To get insight into the significance of $r$, we use the Vieta theorem to rewrite the polynomial in Eq.\ (\ref{SM.quartic3}) as
\begin{align}
E^4 + p E^2 + q E + r = (E-E_1)(E-E_2)(E-E_3)(E-E_4) ,
\end{align}
where $E_i$ are the solutions of the quartic equation. Setting $E=0$ we obtain
\begin{align}
r = E_1 E_2 E_3 E_4 .
\end{align} 
Note that $\det \mathcal{H} = E_1^2 E_2^2 E_3^2 E_4^2 = r^2$ is the square of the Pfaffian $P(\mathbf{k})$ of the BdG Hamiltonian unitarily transformed into antisymmetric form. The Pfaffian determines the $\mathbb{Z}_2$ invariant that protects BFSs in centrosymmetric superconductors \cite{SM.ABT17,SM.BAMT18}. Obviously, nodes of any type, including BFSs, are characterized by zeros of $\det \mathcal{H}$ and thus by zeros of the Pfaffian. At BFSs specifically, the Pfaffian generically changes sign as a function of momentum $\mathbf{k}$. We find that $r$ equals the Pfaffian $P(\mathbf{k})$, up to the sign. However, the overall sign of the Pfaffian is not unitarily invariant and therefore not physically meaningful \cite{SM.ABT17,SM.BAMT18}. We can thus choose this sign in such a way that $r=P(\mathbf{k})$.

Equation (\ref{SM.Pfrl}) can be rewritten as
\begin{align}
r &=  (\langle c, c \rangle - \langle \Delta^1, \Delta^1 \rangle
  - \langle \Delta^2, \Delta^2 \rangle)^2
  + 4 \left( \langle c, \Delta^1 \rangle^2 + \langle \Delta^1, \Delta^2 \rangle^2
  + \langle \Delta^2, c\rangle^2
  - \langle \Delta^1, \Delta^1 \rangle \langle \Delta^2, \Delta^2 \rangle \right) .
\end{align}
We observe that the first term is a complete square. Hence, the Pfaffian $r=P(\mathbf{k})$ can only change sign when both $\vec{\Delta}^1$ and $\vec{\Delta}^2$ have at least one nonzero component.

With the values of $p$, $q$, and $r$ at hand, the solutions of the quartic equation can be obtained in closed form using the Ferrari-Cardano method. However, it turns out to be numerically more robust to obtain them as the eigenvalues of the companion matrix~\cite{SM.HJ12}
\begin{align}
C = \begin{pmatrix}
    0 & 0 & 0 & -r \nonumber \\
    1 & 0 & 0 & -q \nonumber \\
    0 & 1 & 0 & -p \nonumber \\
    0 & 0 & 1 & 0
  \end{pmatrix} .
\end{align}

\subsection{Derivatives of eigenenergies}

The BCS gap equation {(8)} in the main text, as well as the inverse gap equation {(9)}, contain derivatives of eigenenergies with respect to the pairing amplitudes $\Delta_n^\alpha$. For high-precision results, it is beneficial to evaluate these derivatives analytically instead of numerically using a finite-difference formula. Noting that the eigenenergies are solutions of the  quartic equations $E^4 + p E^2 + q E + r = 0$, the derivatives $\partial E_{\mathbf{k},i}/\partial \Delta_n^\alpha$ can be expressed in terms of derivatives of the coefficients $p$, $q$, and $r$:
\begin{equation}
\frac{\partial E}{\partial \Delta_n^\alpha}
  = \frac{\partial E}{\partial p}\, \frac{\partial p}{\partial \Delta_n^\alpha}
  + \frac{\partial E}{\partial q}\, \frac{\partial q}{\partial \Delta_n^\alpha}
  + \frac{\partial E}{\partial r}\, \frac{\partial r}{\partial  \Delta_n^\alpha} ,
\label{SM.dEdD0}
\end{equation}
where
\begin{align}
\frac{\partial E}{\partial p} &= - \frac{E^2}{4 E^3 + 2 p E + q} , \\
\frac{\partial E}{\partial q} &= - \frac{E}{4 E^3 + 2 p E + q} , \\
\frac{\partial E}{\partial r} &= - \frac{1}{4 E^3 + 2 p E + q} .
\end{align}
Inserting these equations into Eq.\ (\ref{SM.dEdD0}) yields
\begin{equation}
\frac{\partial E}{\partial \Delta_n^\alpha} = - \frac{1}{4 E^3 + 2 p E + q}
  \left( E^2\, \frac{\partial p}{\partial \Delta_n^\alpha}
  + E\, \frac{\partial q}{\partial \Delta_n^\alpha}
  + \frac{\partial r}{\partial \Delta_n^\alpha} \right) .
\label{SM.dEdD}
\end{equation}

The closed expressions for the coefficients $p$, $q$, and $r$ derived above allow us to write down the derivatives with respect to $\Delta_n^\alpha$. To start with, Eq.\ (\ref{SM.p}) simply gives
\begin{equation}
\frac{\partial p}{\partial \Delta_n^\alpha} = -4 \Delta_n^\alpha .
\label{SM.dpdD}
\end{equation}
If $q\neq 0$ we can write
\begin{equation}
\frac{\partial q}{\partial \Delta_n^\alpha}
  = \frac{1}{2q}\, \frac{\partial q^2}{\partial \Delta_n^\alpha} .
\label{SM.dqdD}
\end{equation}
Equation (\ref{SM.q}) shows that $q$ does not depend on $\Delta_0^\alpha$. Thus, we obtain $\partial q^2/\partial\Delta_0^\alpha=0$ and
\begin{align}
\frac{\partial q^2}{\partial \Delta_n^\alpha}
  &= 64\, \frac{\partial}{\partial \Delta_n^\alpha}\, \det M \nonumber \\
&= 128\, \Big\{
  (\vec c\cdot\vec c)(\vec \Delta^{\bar\alpha}\cdot\vec \Delta^{\bar\alpha})\, \Delta_n^\alpha
  + (\vec c\cdot\vec \Delta^{\bar\alpha})(\vec \Delta^1\cdot\vec \Delta^2)\, c_n
  + (\vec c\cdot\vec \Delta^1)(\vec c\cdot\vec \Delta^2)\, \Delta_n^{\bar\alpha}
  - (\vec c\cdot\vec \Delta^{\bar\alpha})^2\, \Delta_n^\alpha \nonumber \\
&\quad{}- (\vec c\cdot\vec c)(\vec \Delta^1\cdot\vec \Delta^2)\, \Delta_n^{\bar\alpha}
  - (\vec c\cdot\vec \Delta^\alpha)(\vec \Delta^{\bar\alpha}\cdot\vec \Delta^{\bar\alpha})\,
    c_n \Big\} \nonumber \\
&= 128\, \Big\{
  \big[ (\vec c\cdot\vec \Delta^{\bar\alpha})(\vec \Delta^1\cdot\vec \Delta^2)
    - (\vec c\cdot\vec \Delta^\alpha)(\vec \Delta^{\bar\alpha}\cdot\vec \Delta^{\bar\alpha})
    \big]\, c_n
  + \big[ (\vec c\cdot\vec c)(\vec \Delta^{\bar\alpha}\cdot\vec \Delta^{\bar\alpha})
    - (\vec c\cdot\vec \Delta^{\bar\alpha})^2 \big]\, \Delta_n^\alpha \nonumber \\
&\quad{}+ \big[ (\vec c\cdot\vec \Delta^1)(\vec c\cdot\vec \Delta^2)
    - (\vec c\cdot\vec c)(\vec \Delta^1\cdot\vec \Delta^2) \big]\,
      \Delta_n^{\bar\alpha} \Big\}
\label{SM.dq2dD}
\end{align}
for $n\ge 1$, where $\bar\alpha=2$ ($1$) for $\alpha=1$ ($2$).

Finally, from Eq.\ (\ref{SM.Pfrl}), we obtain
\begin{align}
\frac{\partial r}{\partial \Delta_n^\alpha}
&= 4s_n \langle \Delta^\alpha,\Delta^\alpha\rangle\, \Delta_n^\alpha
  + 8s_n \langle c,\Delta^\alpha\rangle\, c_n
  + 8s_n \langle \Delta^1,\Delta^2\rangle\, \Delta_n^{\bar\alpha}
  - 4s_n \langle c,c\rangle\, \Delta_n^\alpha
  - 4s_n \langle \Delta^{\bar\alpha},\Delta^{\bar\alpha}\rangle\, \Delta_n^\alpha \nonumber \\
&= 4s_n \left[ 2 \langle c,\Delta^\alpha\rangle\, c_n
  + \big( \langle \Delta^\alpha,\Delta^\alpha\rangle
    - \langle \Delta^{\bar\alpha},\Delta^{\bar\alpha}\rangle
    - \langle c,c\rangle \big)\, \Delta_n^\alpha
  + 2 \langle \Delta^1,\Delta^2\rangle\, \Delta_n^{\bar\alpha} \right] ,
\label{SM.drdD}
\end{align}
where $s_n=1$ ($-1$) for $n=0$ ($n\ge 1$). This concludes the derivation of the explicit form of the derivatives appearing in the gap equation. The results show explicitly that all the derivatives are at least of first order in $\Delta_n^\alpha$, which has been used in the derivation of the inverse gap equation in the main text.

\section{Weak-coupling scaling}
\label{SM.scaling}

In this section, we briefly review the scaling of the pairing amplitude $\Delta$ and the free (internal) energy with the interaction strength at temperature $T=0$. By performing the momentum integration in Eq.\ {(6)} in the main text and expanding up to order $\Delta^2$, we obtain the free-energy difference per unit cell as
\begin{equation}
\Delta F = F_s-F_n
  \cong c_1 \Delta^2 \ln \Delta + c_2 \Delta^2 + \frac{1}{2}\, \frac{\Delta^2}{V}
  = \left( c_1 \ln \Delta + c_2 + \frac{1}{2V} \right) \Delta^2 ,
\label{SM.DF.0}
\end{equation}
where $c_1$ and $c_2$ are coefficients and $V$ is the interaction strength, see Eq.\ {(7)} in the main text. The last term stems from the mean-field decoupling in the Cooper channel. The expansion of the momentum sum for small $\Delta$ generates a leading term proportional to $\Delta^2 \ln \Delta$ (the Cooper logarithm), another contribution of order $\Delta^2$, and terms of orders higher than $\Delta^2$, which are omitted here. A normalizing factor inside the logarithm has been absorbed into $c_2$.


The saddle point is found by taking
\begin{equation}
0 \stackrel{!}{=} \frac{\partial\Delta F}{\partial\Delta} \cong c_1 \Delta
  + 2 \left( c_1 \ln \Delta + c_2 + \frac{1}{2V} \right) \Delta
  = \left( c_1 + 2c_2 + \frac{1}{V} + 2 c_1 \ln \Delta \right) \Delta .
\end{equation}
For the nontrivial solution, we find
\begin{equation}
\ln \Delta \cong - \frac{1}{2} - \frac{c_2}{c_1} - \frac{1}{2c_1V}
  \equiv A - \frac{B}{V} .
\label{SM.logDelta}
\end{equation}
This is the weak-coupling scaling form of the pairing amplitude used in the main text.

From Eq.\ (\ref{SM.logDelta}), we also obtain the well-known BCS weak-coupling form of the pairing amplitude,
\begin{equation}
\Delta \cong e^{-1/2-c_2/c_1}\, e^{-1/2c_1V}
\end{equation}
and thus
\begin{equation}
\Delta^2 \cong e^{-1-2c_2/c_1}\, e^{-1/c_1V} .
\label{SM.Delta2}
\end{equation}
Insertion of Eqs.\ (\ref{SM.logDelta}) and (\ref{SM.Delta2}) into Eq.\ (\ref{SM.DF.0}) gives the free energy difference at the saddle point,
\begin{equation}
\Delta F \cong -\frac{c_1}{2}\, e^{-1-2c_2/c_1}\, e^{-1/c_1V} .
\end{equation}
We obtain the scaling form
\begin{equation}
\ln(-\Delta F) \cong \ln\frac{c_1}{2} - 1 - \frac{2c_2}{c_1} - \frac{1}{c_1V}
  \equiv A' - \frac{B'}{V} .
\label{SM.logDF}
\end{equation}
The coefficients in Eqs.\ (\ref{SM.logDelta}) and (\ref{SM.logDF}) are related by
\begin{align}
A' &= \ln\frac{c_1}{2} - 1 - \frac{2c_2}{c_1} = 2A + \ln\frac{c_1}{2}
  = 2A - \ln 4B , \\
B' &= \frac{1}{c_1} = 2B .
\end{align}

\section{Numerical solution of the inverse gap equation}

In this section, we provide some background for the numerical solution of the inverse gap equation. Both the free (internal) energy and the inverse gap equation contain integrations over the three-dimensional Brillouin zone. High precision in these integrations is essential for reaching the weak-coupling regime. Of course, symmetries can be exploited to restrict the integrals to a part of the Brillouin zone but it remains a three-dimensional integral for any lattice model.

The integration is particularly demanding with regard to precision because the weak-coupling behavior relies on a logarithmic term proportional to $\Delta^2 \ln (\Delta/\Lambda)$ in the free (internal) energy at $T=0$, which has to be separated from a $\Delta^2$ contribution, where both are exponentially small for weak pairing interaction. Here, $\Lambda$ is a high-energy cutoff resulting from the finite band width. The Cooper logarithm $\ln (\Delta/\Lambda)$ results from integration over the entire Brillouin zone, as shown by the presence of both the low-energy scale $\Delta$ and the high-energy scale $\Lambda$.

We find that the common method of summing over a momentum-space mesh becomes forbiddingly slow for three-dimensional systems if reasonable accuracy is desired. Instead, we have obtained good results using adaptive integration. We perform the integration using spherical coordinates. The radial integration is performed first, inside the angular integrals. In order to capture the effect of the superconducting gap close to the normal-state Fermi momentum $k_F$, the radial integration is split into four parts over the intervals $[0,k_F-k_1]$, $[k_F-k_1,k_F]$, $[k_F,k_F+k_2]$, and $[k_F+k_2,k_{\text{BZ}}(\theta,\phi)]$, where $k_{\text{BZ}}(\theta,\phi)$ describes the surface of the Brillouin zone in the $(\theta,\phi)$ direction. The widths of the regions, i.e., the values of $k_1$ and $k_2$, are proportional to the gap $\Delta$ at $k_F$ and the constants of proportionality are determined so as to minimize the numerical noise. The integrals are performed using globally adaptive sampling as implemented in Mathematica \cite{SM.mathematica} with the accuracy goal set to $18$ digits and the maximum number of recursions set to $12$ for the first and fourth interval and to $8$ for the second and third interval. We also used globally adaptive sampling for the integration over $\theta$ and $\phi$ with the accuracy goal set to $18$ digits and the maximum number of recursions set to~$4$.

The main diagnostics for the quality of the numerical integration are the following: The pairing interaction and the free-energy gain $\Delta F$ are smooth functions of the gap $\Delta$---recall that we are solving the inverse gap equation. If we push the calculations to smaller $\Delta$ than presented here, we observe step-like behavior typical for round-off error. The results in the range of weak pairing interactions agree with the expected weak-coupling scaling of BCS theory, see Sec.\ \ref{SM.scaling}. Moreover, the scaling of the gap $\Delta$ and of the free-energy gain $\Delta F=F_s-F_n$ are consistent with each other.

\section{$T_{2g}$ pairing states}

In this section, we briefly present results for the $T_{2g}$ pairing states obtained by solving the inverse gap equation. The results are largely analogous to the $E_g$ case. The $T_{2g}$ states are characterized by a three-component order parameter $(\Delta_1,\Delta_2,\Delta_3) \equiv \Delta\, \bdelta$ \cite{SM.BAMT18}. Landau analysis predicts that $\bdelta = (1,0,0)$, $(1,1,1)/\sqrt{3}$, $(1,i,0)/\sqrt{2}$, and $(1, \omega, \omega^2)/\sqrt{3}$ as potentially stable \cite{SM.BWW16,SM.BAMT18}, where $\omega = e^{2 \pi i/3}$. Similar to $E_{g}$ pairing, the phase diagram contains a first-order transition from the TRSB $(1,i,0)$ state at weak coupling to the time-reversal-symmetric $(1,0,0)$ state at strong coupling, as shown in Fig.\ \ref{SM.Fig1}. Also like for $E_g$ pairing, the pairing amplitude $\Delta$ and the free-energy gain $\Delta F$ exhibit weak-coupling scaling for small pairing strength $V_T$ and the numerical method is robust far into the weak-coupling regime. For larger $V_T$, all states show the S shape for $\Delta$ and swallowtail for $\Delta F$ characteristic for a first-order transition.

For $T_{2g}$ pairing, there are two TRSB states. Of these, the chiral $(1,i,0)$ state is weakly favored over the cyclic $(1,\omega,\omega^2)$ state at weak coupling, as shown in the inset of Fig.\ \ref{SM.Fig1}(b). Recall that the largest value corresponds to the stable state since the factor $1/\Delta F_{(1,i,0)}$ is negative. This result is surprising in view of the well-known arguments by Sigrist and Ueda \cite{SM.SiU91}: For infinitesimal coupling, the $(1,i,0)$ state has a line node on the equator of the normal-state Fermi surface and two point nodes at the poles \cite{SM.ABT17,SM.BAMT18}. The $(1,\omega,\omega^2)$ state instead has eight point nodes \cite{SM.BAMT18}. Hence, one would expect the DOS close to the Fermi energy to be lower for the latter state. However, the quasiparticle dispersion close to the point nodes for the $(1,\omega,\omega^2)$ state is rather shallow compared to the $(1,i,0)$ state for otherwise identical parameters. Evidently, this destabilizes the $(1,\omega,\omega^2)$ state also in the limit of infinitesimal coupling. The main insight here is that a TRSB state with a line node can be stabilized over a TRSB state that only has point nodes.

The TRSB $T_{2g}$ states develop BFSs for increasing $V_T$, which destabilize them relative to the time-reversal-symmetric $(1,0,0)$ and $(1,1,1)$ states. Initially, the $(1,0,0)$ state is weakly favored, as shown in the main panel of Fig.\ \ref{SM.Fig1}(b). For larger $V_T$, the free-energy gains of the $(1,0,0)$ and $(1,1,1)$ states approach each other but $(1,0,0)$ seems to remain favored.

\begin{figure}
\raisebox{1ex}{(a)}\includegraphics[width=0.5\textwidth]{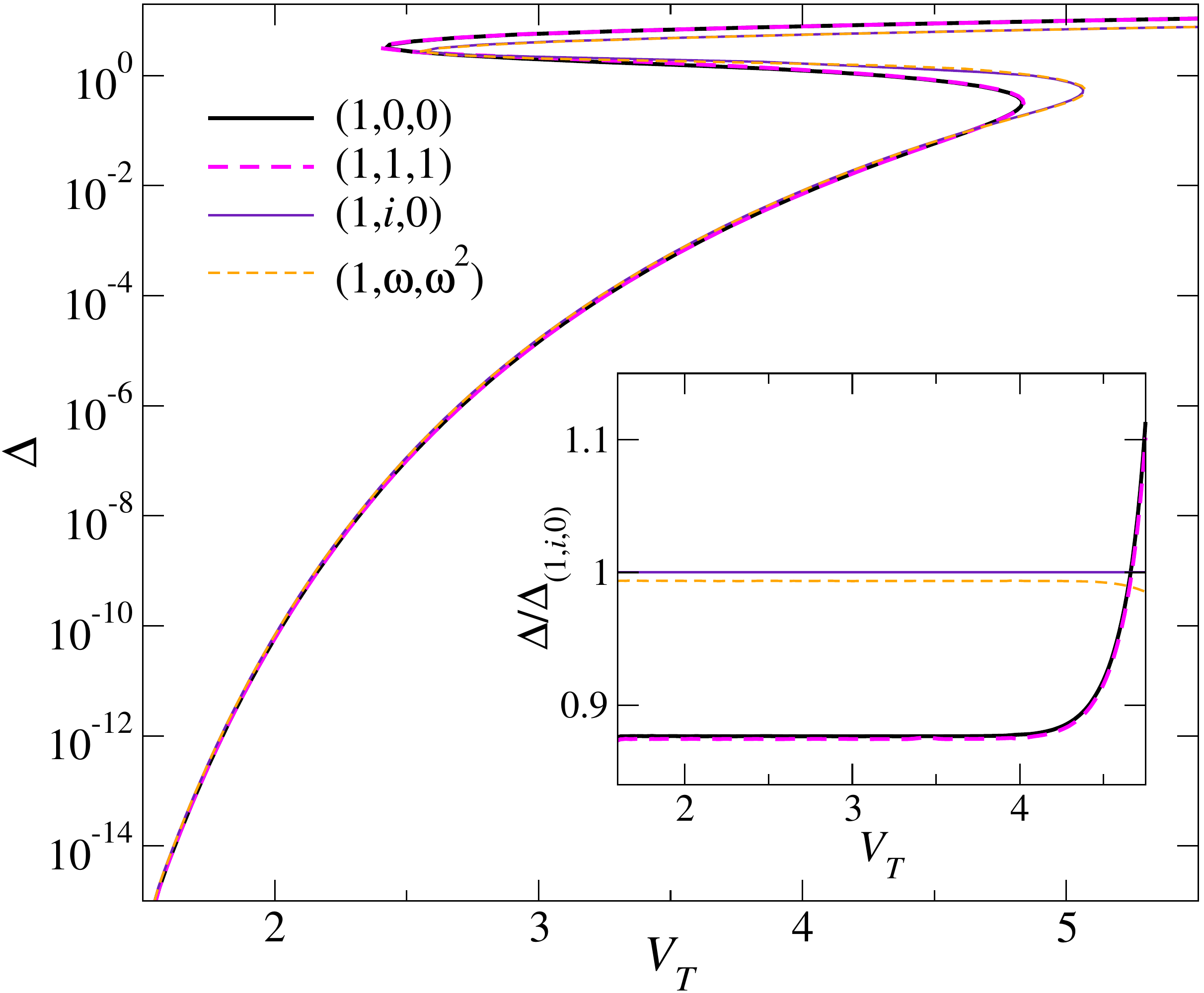}\\[1.5ex]
\raisebox{1.7ex}{(b)}\includegraphics[width=0.5\textwidth]{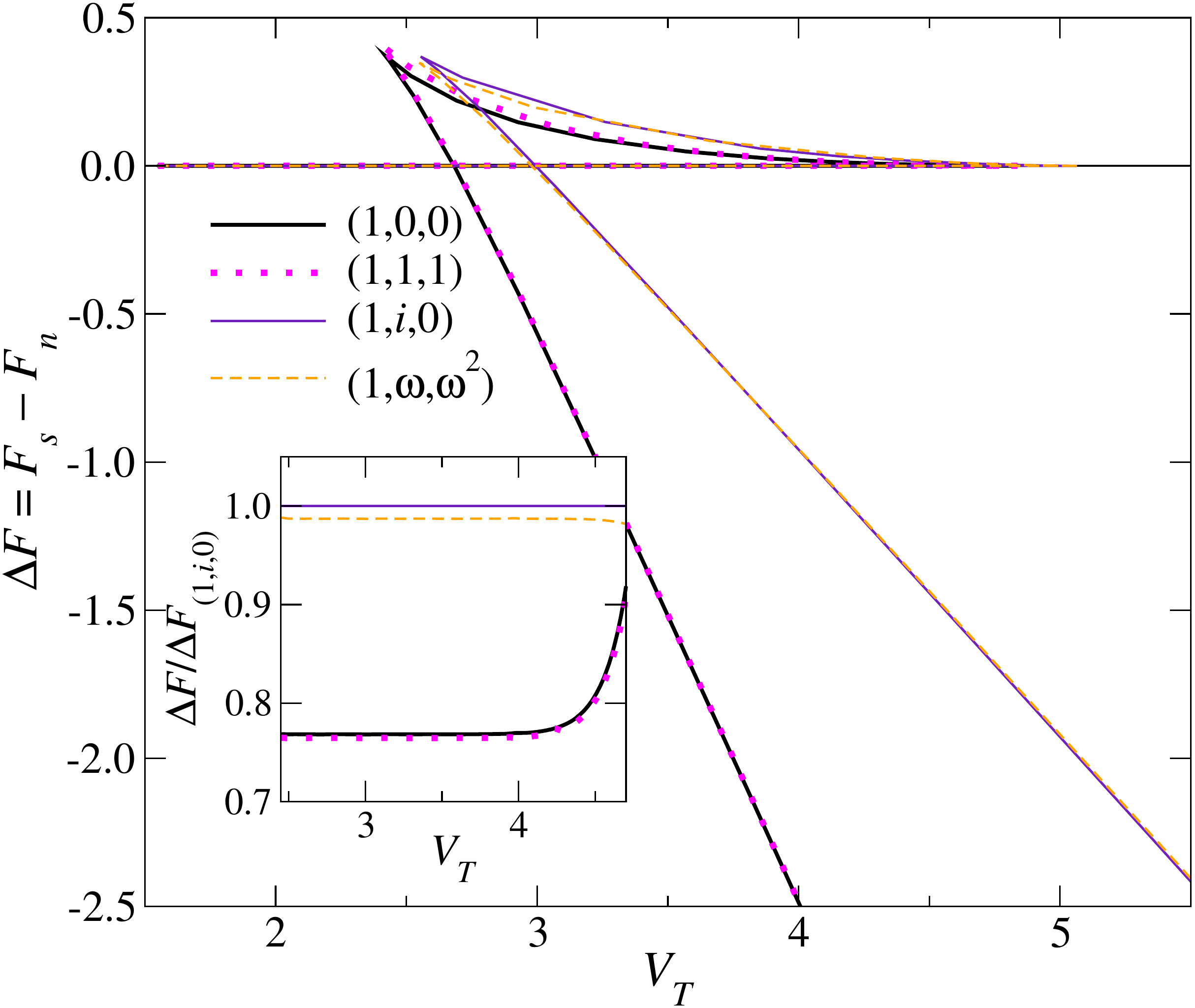}
\caption{(a) Pairing amplitudes $\Delta$ as functions of the coupling strength $V_T$ at $T=0$ for the $T_{2g}$ pairing states $(1,0,0)$, $(1,1,1)$, $(1,i,0)$, and $(1,\omega,\omega^2)$ with $\omega = e^{2 \pi i/3}$. The inset shows the ratios of $\Delta$ for the four states to $\Delta$ for the $(1,i,0)$ state, which is the ground state at weak coupling.
(b) Free-energy differences $\Delta F = F_s-F_n$ between the superconducting and normal states as functions of the coupling strength $V_T$ at $T=0$ for the same $T_{2g}$ pairing states. The inset shows the ratios of $\Delta F$ for the three states to $\Delta F$ for the $(1,i)$ state. Details of the models and the numerical parameters are given in Sec.~\ref{SM.sec.model}.}
\end{figure}

In this context, it is also worth pointing out that the $T_{2g}$ pairing states require a significantly larger interaction strength than the $E_g$ pairing states to be stabilized. This is clearly seen in Fig.\ {2} in the main text. Since the pairing strengths in the $T_{2g}$ and $E_g$ channels become equal in the spherical limit this suggests that the $E_g$ states, in particular the weak-couping $(1,i)$ state, are favored for systems with small normal-state Fermi surface. A different scenario that favors $E_g$ pairing over both $T_{2g}$ and conventional $A_{1g}$ pairing has been proposed in Ref.~\cite{SM.KBT22} for pyrochlore materials.

\end{document}